# Unveiling the role of Co-O-Mg bond in magnetic anisotropy of Pt/Co/MgO using atomically controlled deposition and *in-situ* electrical measurement

Yumeng Yang[1], Jiaren Yuan[2,3], Long Qi[1], Ying Wang[1], Yanjun Xu[1,4], Xiaowei Wang[1], Yuanping Feng[3], Baoxi Xu[4], Lei Shen[5,a)] and Yihong Wu[1,b)]

[1] *Department of Electrical and Computer Engineering, National University of Singapore, 4 Engineering Drive 3, Singapore 117583, Singapore*

[2] *College of Science, Nanjing University of Aeronautics and Astronautics, Nanjing 210016, People's Republic of China*

[3] *Department of Physics & Centre for Advanced Two-dimensional Materials, National University of Singapore, Singapore 117542, Singapore*

[4] *Data Storage Institute, A\*STAR (Agency for Science, Technology and Research), 2 Fusionopolis Way, 08-01 Innovis, Singapore 138634, Singapore*

[5] *Department of Mechanical Engineering & Engineering Science Programme, Faculty of Engineering, National University of Singapore, Singapore 117575, Singapore*

Despite the crucial role of interfacial perpendicular magnetic anisotropy in Co(Fe)/MgO based magnetic tunnel junction, the underlying mechanism is still being debated. Here, we report an anatomical study of oxygen and Mg effect on Pt/Co bilayers through repeated *in-situ* anomalous Hall effect measurements, controlled oxygen exposure and Mg deposition in an ultrahigh vacuum system. We found that chemisorbed oxygen not only quenches the effective magnetic moment of the Co surface layer, but also softens its magnetic anisotropy. However, a subsequent Mg dusting on the oxygen pre-exposed Pt/Co surface can recover the magnetic anisotropy. The *ab initio* calculations on the exchange splitting and orbital hybridization near the Fermi level give a clear physical explanation of the experimental observations. Our results suggest that Co(Fe)-O-M bond plays a more important role than the widely perceived Co(Fe)-O bond does in realizing interfacial perpendicular magnetic anisotropy in Co(Fe)/MgO heterostructures.

# I. INTRODUCTION

Perpendicular magnetic anisotropy (PMA) is of great importance in building spin transfer torque (STT) or spin orbit torque (SOT) based spintronic devices, due to its reduced process variation and excellent down scaling capability [1,2]. One class of materials that simultaneously exhibits stable PMA, large tunneling magnetoresistance (TMR), and low switching current is the ferromagnetic (FM) transition metal/oxide heterostructure, including CoFeB/MO$_x$ and Co(Fe)/MO$_x$ (M = Mg, Al, *etc.*) [3,4]. Experimentally it has been well established that a large PMA [5-10] or TMR [11-14] can be obtained through either optimized plasma (natural) oxidation [5,6,11-13] or post-annealing [9,10] or the combination of both processes [7,8,14]. Results from *ab initio* calculations [15-17] suggest that the PMA originates from strong hybridization between (Co)Fe-3$d$ and the O-2$p$ orbitals at the (Co)Fe/oxide interface, which can penetrate into (Co)Fe for a few atomic layers. It was further pointed out that an abrupt interface with Co(Fe)-O bond is more desirable for a sizable PMA than either the under-oxidized interface with Co(Fe)-M bond or over-oxidized one with Co(Fe)-O-Co(Fe) bond. The importance of Co(Fe)-O bond for obtaining the PMA has recently been confirmed experimentally in CoFeB-MgO MTJs by directly imaging the atoms using advanced electron microscopy, wherein it was found that CoFe bonds atomically to MgO grains in an epitaxial manner by forming Co(Fe)-O bond at the interfaces without the incorporation of Co(Fe) into MgO or *vice versa* [18]. In another attempt, a maximum magnetic anisotropy of ~60 meV was obtained by directly placing a single Co atom atop the O site of an MgO (100) surface [19]. Both findings provided direct microscopic evidences that the origin of PMA lies in interfacial Co(Fe)-O bond. However, in all these studies the oxygen comes from the oxide instead of free oxygen atoms or molecules, it thus remains unclear whether Co(Fe)-O alone or Co(Fe)-O-M bond plays a more important role in forming the PMA. In order to elucidate the respective roles of Co-O and Co-O-Mg bonds in forming the PMA, we systematically studied the oxygen and Mg effect on the anisotropy of Pt/Co bilayers using an ultrahigh vacuum (UHV) system with a base pressure

< $5\times10^{-9}$ mbar, which allows to perform *in-situ* deposition of Co and Mg with atomic layer accuracy, controlled adsorption of oxygen and anomalous Hall effect (AHE) measurements without breaking the vacuum.

The experiment began with the deposition of a Pt underlayer which is used to induce PMA in the subsequently deposited Co. In order to study the respective role of oxygen and Mg, we performed O exposure and Mg deposition in sequences between which electrical measurements were carried out. This is in contrast with previous studies in which O and Mg are deposited simultaneously in the form of MgO [9,20]. By gradually increasing the Co thicknesses ($t_{Co}$) in Pt/Co bilayers, we observed the onset of PMA at $t_{Co} \approx 0.6$ nm, and spin reorientation transition (SRT) [21-23] beyond $t_{Co} \approx 1.7$ nm, in which the easy axis changes from perpendicular to in-plane direction. Depending on $t_{Co}$, the subsequent oxygen exposure has a different effect on the effective anisotropy: it softens the PMA at $t_{Co} = 0.6$ and 0.8 nm, but it enhances the PMA at $t_{Co} = 1.9$ nm. Both can be understood as mainly caused by the O adsorption induced decrease of the effective magnetic layer thickness of Co, which is different from the reported oxygen effect on PMA of Co(Fe) in Co(Fe)/MO$_x$ heterostructures [6,7,18]. Further deposition of an Mg dusting layer on top of the oxygen pre-exposed Pt/Co surface recovers the PMA, while direct dusting of Mg on clean Pt/Co surface reduces the PMA. Using a*b initio* calculations, we found that upon O adsorption the reduction of exchange splitting by charge transfer quenches the moment of the topmost Co layer, whereas the subsequently adsorbed Mg adatoms recover the moment by transferring some electrons back to Co. The rebalancing of charge transfer recovers the PMA. Although the sample structure under investigation is not exactly the same as the widely studied CoFeB/MgO system, our combined experimental and theoretical studies implies that Co(Fe)-O-Mg bond plays an important role in the realization of strong PMA in Co(Fe)/oxide heterostructures instead of the widely perceived Co-O bond only.

The remainder of this paper is organized as follows. Section II describes the experimental details. Section III A presents the Co thickness dependence of the effective magnetic anisotropy. In Section III B, we presents the experimental results of oxygen and Mg effect on the magnetic anisotropy. The *ab initio* calculation results are presented and discussed in Section III C to elucidate the roles of each element in Co-O-Mg bond, followed by conclusions in Section IV.

## II. EXPERIMENTAL DETAILS AND THEEORETICAL CALCULATIONS

All the samples were deposited on Si/SiO$_2$ (300 nm) substrate, which was cut into 2 mm × 2 mm squares in order to accommodate the small gap of the electromagnet inside the UHV chamber which provides a perpendicular field up to 2 kOe. In order to avoid short circuit by Pt deposited on the sample holder, an underlayer of Ta(1.5)/Pt(3) (unless specified otherwise and the number inside the parentheses indicates the thickness in nm) was deposited *ex-situ* in a DC magnetron sputter with a base pressure < 4×10$^{-8}$ mbar and process pressure of 4×10$^{-3}$ mbar. After the deposition of Ta/Pt, the sample was immediately loaded into the Omicron UHV system and annealed at 110 °C for 1 h to remove moisture on the surface before subsequent *in-situ* deposition and electrical measurements. Details about the UHV system can be found elsewhere [24-26]. Oxygen exposure and AHE measurements were performed in the same chamber. The oxygen exposure was carried out by gradually increasing the partial pressure from the background vacuum to 1×10$^{-8}$ – 3×10$^{-7}$ mbar through controlling the duration between 5 – 60 minutes. The AHE measurements were performed by directly probing the four corners of the sample. The deposition of Co and Mg was done *in-situ* using K-cells in the preparation chamber, with a rate of 0.033 Å s$^{-1}$ and 0.265 Å s$^{-1}$, respectively. As we will discuss shortly, in the Mg dusting experiment, in order to minimize the amount of Mg deposited on the Co surface, we heated up the Mg source but kept the shutter closed. Mg dusting of Co surface was achieved through Mg atoms leaked out from the K-cell through the small spacing between the shutter and the cell. There was no exposure to ambient after the sample was loaded into the UHV system and throughout the *in-situ* studies.

Density functional theory (DFT) with the projected augmented wave (PAW) method was performed by employing the *Vienna ab initio Simulation Package (VASP)* [27]. Generalized gradient approximation (GGA) in the form of Perdew-Burke-Ernzerhof (PBE) [28] was chosen as exchange correlation potential. The electron wave functions were expanded by a plane wave basis set with a cutoff energy of 500 eV. The Brillouin zone is sampled by a 21×21×1 *k*-point mesh and the convergence criterion of total energy is set to be $10^{-6}$ eV. In general, the calculation includes three steps. First, full structural optimizations were performed until none of the forces exceeded 0.01 eV Å$^{-1}$. Next, the Kohn-Sham equations were solved with a collinear calculation without spin-orbit coupling (SOC) to determine the charge distribution of the system ground state. Finally, SOC was taken into account and the magnetic anisotropy energy (MAE) was calculated as $E_{MAE} = -\frac{1}{a^2}(E_\perp - E_\parallel)$, where *a* is the in-plane lattice constant, and $E_\perp$ ($E_\parallel$) the total energy of the system with spins oriented in out-of-plane (in-plane) direction. As will be discussed shortly, at the Co thickness range investigated, the contribution to effective anisotropy from the bulk magnetocrystalline anisotropy is much smaller than interface anisotropy ($K_s$), and therefore $E_{MAE}$ can be approximated as equal to $K_s$ in the macro-spin model.

## III. RESULTS AND DISCUSSION

### A. Co thickness dependence of magnetic anisotropy revealed by AHE measurements

In the first round of measurements, $t_{Co}$ was systematically varied to understand how the effective magnetic anisotropy of Co/Pt bilayer depends on $t_{Co}$. Figs. 1(a), 1(b) and 1(c) show the AHE loops for Ta(1.5)/Pt(3)/Co($t_{Co}$) with $t_{Co}$ = 0.2 – 1.9 nm. Due to the large range of sweeping field needed for samples with different Co thicknesses and for the sake of clarity, we present the results in three separate sub-plots: (a) $t_{Co}$ = 0.2 nm and 0.4 nm, (b) $t_{Co}$ = 0.6 – 1.7 nm, and (c) $t_{Co}$ = 1.8 nm and 1.9 nm. The offset of AHE raw data was corrected and the curves are vertically shifted for clarity except for the first curve in each sub-plot. As can be seen in Fig. 1(a), there is hardly any observable AHE at $t_{Co}$ = 0.2 nm, presumably caused by the non-conformal coverage of Pt by Co at this small thickness. When Co forms discrete islands on Pt, due to size-effect, its crystalline magnetic anisotropy may become too small to support ferromagnetism at room temperature; this may explain why the AHE signal is diminished $t_{Co}$ = 0.2 nm. With increasing the thickness, the Co islands will grow and coalesce to form large patches, leading to increase in magnetic anisotropy. As not the entire film exhibits PMA, the AHE curve has a mixture of characteristics of PMA and in-plane magnetic anisotropy (IMA) films, as shown in Fig. 1(a) for $t_{Co}$ = 0.4. When $t_{Co}$ is further increased to 0.6 nm, as shown in Fig. 1(b), nearly a square shaped loop was observed, suggesting the onset of PMA. The squareness of the AHE curves remains almost the same until $t_{Co}$ reaches 1.6 nm, though the coercivity of Co increases significantly from 0.6 nm to 1.2 nm and then gradually decreases. At $t_{Co}$ = 1.7 nm, part of the film starts to exhibit IMA, which becomes more dominant over PMA when $t_{Co}$ increases to 1.8 and 1.9 nm, as shown in Fig. 1(c).

The transition from PMA to IMA in Fig. 1(c) is similar to the typical SRT behavior observed in ultra-thin Co or Fe films [22,23]. The transition can be understood as a continuous reorientation of the easy axis from the perpendicular to in-plane direction, via the intermediate "easy cone state" with easy axis canted from the perpendicular direction [29-31]. Since the field strength of electromagnet in our UHV chamber is insufficient for quantifying PMA by performing Hall measurements using an in-plane field, we employ the macro-spin approximation to gain some insights on PMA through fitting the AHE curves at different Co thicknesses near the SRT region. Following the coordinate notion in Fig. 2(a) and with the applied field in z-direction ($\theta_H$ = 0), the free energy density of the film can be expressed as [29]:

$$E = K_{eff} \sin^2 \theta + K_2 \sin^4 \theta - HM_s \cos \theta \qquad (1)$$

where $K_{eff}$ is the effective anisotropy constant defined phenomenologically as $K_{eff} = K_1 - 2\pi M_s^2 + K_s/t_{Co}$ with $K_s$ the interface anisotropy constant, $K_1$ and $K_2$ the second and forth order magnetocrystalline anisotropy constant, $M_s$ the saturation magnetization, $H$ the applied magnetic field, $\theta$ the angles between magnetization and z-direction. For every set of $K_1$, $K_2$, $K_s$, and $M_s$ values, the equilibrium magnetization direction $\theta(H, t_{Co})$ can be obtained numerically through energy minimization, from which the AHE curve can be obtained.

Before proceeding to numerical calculations, it is useful to estimate the range of $K_1$, $K_2$ and $K_s$ analytically. The energy minimization requires $\frac{\partial E}{\partial \theta} = 0$, which yields:

$$2\sin 2\theta (K_{eff} + 2K_2 \sin^2 \theta + \frac{HM_s}{2\cos\theta}) = 0 \qquad (2)$$

Eq. (2) has two sets of solutions. By correlating the experimental data in Figs. 1(a) – 1(c) with the three anisotropy cases discussed in the Appendix, it can be identified that the slanted loops at $t_{Co}$ = 1.8 nm and 1.9 nm are associated with the solutions of $K_{eff} + 2K_2 \sin^2 \theta + \frac{HM_s}{2\cos\theta} = 0$, whereas the relatively square loops at $t_{Co}$ = 1.6 nm and 1.7 nm come from the solution of $2\sin 2\theta = 0$. For the former case, we further plotted $-HM_s/(2\cos\theta)$ against $2\sin^2\theta$ with $M_s = 1407$ emu cm$^{-3}$ for the two samples in Fig. 2(b). In the plot, $\cos\theta$ is calculated by normalizing $R_{xy}$ at each $H$ and only the data from the reversible portion below saturation field is used. In this way, $K_{eff}$ and $K_2$ at $t_{Co}$ = 1.8 nm and 1.9 nm can be estimated by a linear fitting. For the latter case, the magnetization switches at $H = \pm\frac{2K_{eff}}{M_s}$; therefore, $K_{eff}$ at $t_{Co}$ = 1.6, 1.7 nm can be obtained from the $H_c$ of the AHE curve. In Fig. 2(c), $K_{eff}$ values at different thicknesses are plotted against $1/t_{Co}$. Through linear fitting, $K_1$ and $K_s$ are estimated as $4.47\times10^6$ erg cm$^{-3}$ and 1.33 erg cm$^{-2}$, respectively.

We now turn to numerical minimization of Eq. (1) by using the obtained $K_1$, $K_2$ and $K_s$ as the starting values. Due to the sensitivity of AHE curves to thickness at around the SRT critical thickness, it is difficult to fit the curves by assuming a uniform $t_{Co}$ across the entire sample. In reality, it is very likely that the sample consists of a mixture of PMA, "easy cone" and IMA states due to subtle thickness variation over a relatively large size sample. To account for the thickness effect, we assumed that the sample consists of areas with different Co thickness and the partial area of the film at thickness $t_{Co}$ follows a normal distribution $f(t_{Co}, \bar{t}_{Co}) = \frac{1}{\sqrt{2\sigma^2 \pi}} \exp\left[-\frac{(t_{Co} - \bar{t}_{Co})^2}{2\sigma^2}\right]$ with $\bar{t}_{Co}$ the average thickness, $\sigma$ the standard deviation, and $t_{Co}$ taken in the range of $0 - 2\bar{t}_{Co}$. With this assumption the AHE loop at $\bar{t}_{Co}$ is calculated as $R_{xy}(H, \bar{t}_{Co}) = \int_0^{2\bar{t}_{Co}} f(t_{Co}, \bar{t}_{Co}) \cos\theta(H, t_{Co}) dt_{Co}$. The integration is performed numerically by dividing the Co thickness in the range of $0 - 2\bar{t}_{Co}$ in 1001 steps. As shown in Figs. 2(d) and 1(e), by fitting the AHE loops at $t_{Co} = 1.7 - 1.9$ nm (average nominal thickness of Co), we obtained $K_1 = (4.52 \pm 0.33) \times 10^6$ erg cm$^{-3}$, $K_2 = (1.27 \pm 0.08) \times 10^5$ erg cm$^{-3}$ and $K_s = 1.33$ erg cm$^{-2}$. These results imply that at the present $t_{Co}$ range, $K_{eff}$ is dominated by the $K_s/t_{Co}$ term, and the magnitude of $K_s$ is comparable to the value reported in Pt/Co multilayers (around $0.20 - 1.15$ erg cm$^{-2}$) [32], and that in Pt/Co/AlO$_x$ heterostructures (around $0.64 - 1.74$ erg cm$^{-2}$, calculated using bulk $K_1$ value) [6,7]. On the other hand, for the $t_{Co} = 1.2 - 1.6$ nm samples, as shown in Fig. 2(f), the AHE loops can be fitted without the additional consideration of thickness distribution. The reason for this is that below the critical thickness of around 1.7 nm, despite thickness variations, the entre sample is mostly in the PMA state, and thus the AHE loops remain square shaped. Table I summarizes the parameters used for the fitting of samples with $t_{Co} = 1.2 - 1.9$ nm. It should be noted that $K_1$, $K_2$, $K_s$ values remain almost the same for $t_{Co} = 1.6 - 1.9$ nm, while a much smaller $K_1$ is needed for $t_{Co} = 1.2 - 1.4$ nm. When $t_{Co}$ is below 1.0 nm, the AHE loops can only be reproduced by using a negative $K_1$ (not shown here), which contradicts the assumption of $K_1 > 0$. In fact, these results can be anticipated from the limitation of the macro-spin model for samples at PMA states, especially with very small thickness. Nevertheless, it is safe to say that the model is suitable to account for most of the experimental observations near SRT thickness region.

**B. Oxygen exposure and Mg dusting effect on magnetic anisotropy**

Next we present the oxygen effect on the magnetic anisotropy of Pt/Co bilayers. Figs. 3(a) and 3(b) show the AHE loops for another sample with a structure of Ta(1.5)/Pt(3)/Co(0.6) at different oxygen exposure doses (here L is the Langmuir unit with one Langmuir corresponding to an exposure of $1.33\times10^{-6}$ mbar for one second). A full coverage of the energy favorable *fcc* hollow sites on *hcp* Co surface (to be discussed later) thus requires an oxygen exposure dose of about 5.2 L, assuming a unity sticking coefficient. Again all the curves but the lowest one (without O exposure) are vertically shifted. As can be seen, both the coercive field and AHE signal decrease as the dose increases, both of which are signatures of the gradual transition from PMA to IMA. Moreover, as shown in Fig. 3(a), the exposed sample can return to the original PMA state after a mild annealing at 110 °C for 1 hour in UHV. At this stage, one may be tempted to associate the transition to O exposure induced decrease in $K_s$, which is indeed the case as revealed by *ab initio* calculations (to be discussed later) for ultrathin Co layer. However, as shown in Fig. 3(c) for a thicker sample with the structure of Ta(1.5)/Pt(3)/Co(1.9), upon O exposure, both the coercive field and AHE signal show the opposite trend, *i.e.,* the transition from IMA to PMA. Similarly, the AHE loop recovers after a mild annealing. The different oxygen dose dependence in Fig. 3 for the two samples suggests at least that the oxygen exposure effect cannot be explained by the change in $K_s$ alone. Instead, the behavior of both samples can be explained reasonably well by taking into the additional consideration that the oxygen exposure seems to induce an effect which corresponds to an effective reduction of $t_{Co}$. To shed more light on this point, Fig. 4(a) compares the AHE loops for two sets of samples: i) a pristine sample with $t_{Co} = 0.4$ nm and an O exposed sample with $t_{Co} = 0.6$ nm and ii) a pristine sample with $t_{Co} = 1.8$ nm and an O exposed sample with $t_{Co} = 1.9$ nm.

The similarity of AHE loops in both sets of samples agrees with the above hypothesis, *i.e.*, upon O exposure, the effective Co thickness decreases. In fact, it has been reported earlier that the oxygen exposure of transition FMs can cause the partial or complete quenching of magnetism due to either the chemisorption of oxygen or the formation of oxides or both [33,34]. For a more quantitative understanding of the present case, the remanent magnetization ($M_r$) to $M_s$ ratio is extracted from AHE loops at different thicknesses and oxygen doses by assuming $M_r / M_s = R_{xy}(0) / R_{xy}(H_{max})$, where $R_{xy}(0)$, $R_{xy}(H_{max})$ is the Hall resistance at zero field and maximum field, respectively. Figs. 4(b) and 4(c) summarize the ratios at different oxygen doses for the $t_{Co}$ = 0.6 nm exposed sample and the $t_{Co}$ = 1.9 nm exposed sample, respectively. In addition, the ratios of the $t_{Co}$ = 0.4, 1.7, 1.8 nm pristine samples are added in the figures as references. By comparing these results, it can be estimated that the O exposure induced Co thickness reduction is around 0.1 - 0.2 nm under the present exposure conditions. This together with the observation of the recovery of $K_{eff}$ after mild annealing elucidates the main picture of O exposure, *i.e.*, the moment of the topmost Co layer is largely quenched by oxygen adsorption at the Co surface, which in turn induces either PMA→IMA or IMA→PMA transition at small or large Co thickness, respectively. Similar trends were also observed in a few more samples with different $t_{Co}$ values (not shown here). It should be noted that based on the experimental data presented so far, the contribution of Co-O interface to overall $K_s$ including the Pt-Co interface cannot be quantified because its effect is masked out by the more dominant change caused by $t_{Co}$.

The aforementioned dependence of PMA on O dose and decrease of effective $t_{Co}$ upon O exposure are apparently different from the situation in Pt/Co/AlO$_x$ heterostructures [6,7], where a maximum PMA is usually obtained upon oxidation under optimal conditions. This naturally leads to the question about the role of M-O bond in promoting the PMA. To elucidate the role of M in M-O bond, two more sets of experiment were carried out by using Mg as the dusting layer. It should be noted that a relatively thick layer of Mg (~ 0.8 nm) would lead to IMA of Co regardless of whether the Co surface is oxygen exposed or not. If we had used the pre-calibrated deposition rate of 0.265 Å s$^{-1}$, the shutter could only be opened for a few seconds, which would make it difficult to achieve a precise control of the amount of Mg deposited on the Co surface due to manual operation of the shutter. Therefore, in the Mg dusting experiment, we heated up the K-cell to have a nominal deposition rate of 0.265 Å s$^{-1}$, but with K-cell shutter closed during deposition. In the first series of experiments, Mg was dusted on O pre-exposed samples and subsequently exposed the Mg dusted samples to oxygen again (hereafter we refer it as "re-exposed" sample). As an example, Fig. 4(d) compares the AHE loops for the sample Ta(1.5)/Pt(3)/Co(0.8) at different stages including pristine, O pre-exposed (dose of 1218 L), Mg dusted and O re-exposed (dose of 3760 L) states. As can be seen, after Mg dusting and O re-exposure, the sample is almost recovered back to the pristine state. In some other samples (not shown here), PMA was recovered after Mg-dusting without further re-exposure to oxygen, which is probably due to the variation in amount of adsorbed oxygen and/or Mg adatoms in the pre-exposure and/or dusting process among different samples. On the other hand, as shown in Fig. 4(e), direct Mg dusting on a pristine Ta(1.5)/Pt(3)/Co(0.6) sample slightly weakens the PMA, and re-exposure to oxygen (dose of 3548 L) has little effect on it. Since oxygen mainly affects the surface layer, the presence of Mg layer largely protects the Co layer from interacting with oxygen. The weakening of PMA in this case is therefore mostly results from the Co-Mg bond. Both observations in Figs. 4(d) and 4(e) suggest that Mg in MgO indeed plays an active role in Pt/Co/MgO heterostructures.

## C. *Ab initio* calculations of oxygen adsorption and Mg dusting effect

To shed light on the respective roles of each element in Co-O-Mg bond, first-principles calculations are carried out. The pristine Pt/Co is explored firstly as a reference. The schematics of the optimized structures after oxygen exposure (Pt/Co/O) and Mg dusting (Pt/Co/O/Mg) are shown in Figs. 5(a) and 5(b), respectively. Previous studies report that O atoms, not O molecules, are chemisorbed at low dose upon exposure of Co surface because the strong surface interaction can break the O-O bond [35-38]. Our total-energy calculations suggest both adsorbed O and Mg favor the *fcc* hollow sites on the Co surface, which is consistent with low-energy electron diffraction (LEED) studies [39,40] and DFT calculations [41,42]. Table II summarizes the calculated spin moments ($m_S$), orbital moments ($m_L$), and $K_s$ for all the structures with geometry optimization. Notably, spin moment ($m_S$) of the topmost Co layer decreases from 1.82 $\mu_B$ per atom to 0.27 $\mu_B$ per atom upon O adsorption, and recovers to 1.92 $\mu_B$ per atom after adding Mg adatoms. This variation of magnetic moments for topmost Co atom (from unpaired electrons) is directly correlated to the induced change in the charge distribution in the two structures. For a clear view of the charge transfer effect, we depict the charge density difference in Fig. 5(c) for Pt/Co/O and Fig. 5(d) for Pt/Co/O/Mg. As can be seen from the color [see caption of Fig. 5], in Pt/Co/O, the charge transfers from the topmost Co to O atoms due to the high electronegativity of the O atom. Whereas in Pt/Co/O/Mg, O atoms gain electrons directly from Mg atoms, and this in turn results in the transferring of electrons back to Co atoms. More quantitatively, Bader charge analysis shows that the topmost Co layer transfers 0.84 $e^-$ per atom to the O atom in Pt/Co/O, and this electron loss is compensated by Mg with 0.40 $e^-$ per atom transferring back to Co in Pt/Co/O/Mg. In addition, orbital moment ($m_L$) of Co layers follows the same trend as $m_S$, resulting in the variation of $K_s$ [43-46]. It decreases from 0.71 erg cm$^{-2}$ to 0.36 erg cm$^{-2}$ in Pt/Co/O, and returns to 0.70 erg cm$^{-2}$ in Pt/Co/O/Mg. All these calculation results are in qualitative agreement with the experiment ones, although in the experimental case, it is difficult to separate the contribution to effective anisotropy by $K_s$ and the demagnetizing energy.

To have a better understanding of the physical origin of the changes in $m_S$ and $K_s$, the projected densities of the states (PDOS) of *d* orbitals of topmost Co layer and *p* orbitals of O atom are plotted for the cases of Pt/Co/O [Figs. 5(e) and 5(f)] and Pt/Co/O/Mg [Fig. 5(g) and 5(h)]. The PDOS of the reference (Pt/Co) is inserted in 5(h) for comparison. As can be seen, the exchange splitting energy ($E_{\text{exch}}$) in Fig. 5(f) (0.85 eV) is much smaller than that in Fig. 5(h) (2.03 eV) and inset of Fig. 5(h) (2.13 eV), resulting in a reduced value of $m_S$ in Pt/Co/O compared with Pt/Co/O/Mg and Pt/Co. On the other hand, the change in PMA can be understood from the second order perturbation theory, in which $K_s$ is expressed as [47,48]:

$$K_s \propto \xi^2 \sum_k \sum_{o,u} \frac{|\langle k_o | L_z | k_u \rangle|^2 - |\langle k_o | L_x | k_u \rangle|^2}{\varepsilon_{k_u} - \varepsilon_{k_o}} \tag{3}$$

where $\xi$ is an average of the spin orbit coupling (SOC) coefficient, $k_o$ and $k_u$ the occupied and unoccupied states with the wave vector $k$, $L_z$ and $L_x$ the angular momentum operators along $z$ and $x$ directions, respectively, and $\varepsilon_{k_u}$ and $\varepsilon_{k_o}$ the energy of occupied and unoccupied states, respectively. As can be seen from Eq. (3), the SOC between the occupied and unoccupied states with the same magnetic quantum number ($m$) through the $L_z$ operator enhances $K_s$, while that with different $m$ through the $L_x$ operator weakens it. From the PDOS in Fig. 5, we can find that O-$p_x$ and O-$p_y$ ($m = \pm 1$) are degenerate and five $d$ states can be subdivided into $\triangle_1$ ($d_{z^2}$)($m = 0$), $\triangle_3$ ($d_{xz}$, $d_{yz}$)($m = \pm 1$) and $\triangle_4$ ($d_{x^2-y^2}$, $d_{xy}$)($m = \pm 2$) groups. Figs. 5(e) and 5(f) show the hybridization between $p$ states of O atom and $d$ states of Co atom for both occupied and unoccupied states near the Fermi level. Based on Eq. (3), two hybridizations, $\langle p_x | L_x | d_{x^2-y^2} \rangle$ and $\langle p_x | L_x | d_{xy} \rangle$, contribute negatively to the PMA, favoring in-plane anisotropy, while only one orbital hybridization, $\langle p_z | L_z | d_{z^2} \rangle$, contributes positively to PMA. Overall, the adsorbed O leads to the decrease of $K_s$. This is consistent with our experimental result ($t_{Co} = 0.6$ nm) and the previous experimental report that a negative contribution from Co-O interface (hollow sites) $K_s^{Co-O} = -0.04$ erg cm$^{-2}$ was experimentally extracted on hydroxide modified Au(111)/Co surfaces [49]. After depositing Mg, Fig. 5(h) shows that O states near the Fermi level are significantly reduced, resulting in the recovery of $K_s$. We did not observe a notable enhancement of PMA in the Co-O-Mg interfaces as compared to Pt/Co, which is presumably caused by the fact that, in this case, the PMA from Pt-Co is more dominant.

## IV. CONCLUSIONS

In summary, we have performed an anatomical study of O and Mg effect on the magnetic anisotropy of Pt/Co bilayers using *in-situ* AHE measurements. It was found that the oxygen adsorption affects the effective magnetic thickness of Co and thereby changes its magnetic anisotropy. The subsequent Mg dusting can recover the magnetic moment as well as the magnetic anisotropy. *Ab initio* calculations unveil the underlying physics of the change of magnetic moment and interfacial PMA. Our results suggest that the role of Co-M-O bond in the realization of PMA at (Co)Fe/$MO_x$ interfaces may have been overlooked in previous studies relative to the Co(Fe)-O bond. Our work may stimulate further studies on this important interface by adopting a more holistic approach.

## ACKNOWLEDGMENTS

The authors thank Ms. Ziyan Luo for her help in sample preparation and Dr. Xian Qin for her help in discussion and calculation. Y.H.W. would like to acknowledge the support by Ministry of Education, Singapore under its Tier 2 Grant (Grant No. MOE2013-T2-2-096) and the Singapore National Research Foundation, Prime Minister's Office, under its Competitive Research Programme (Grant No. NRF-CRP10-2012-03). L.S. would like to acknowledge the support by the AcRF Tier 1 Research Project (R-144-000-361-112). Y.H.W., Y.P.F. and L.S. are members of the Singapore Spintronics Consortium (SG-SPIN).

## APPENDIX: MACRO-SPIN MODEL OF BILAYER

The Pt/Co bilayer can be phenomenologically treated using the macro-spin model, in which the magnetization vector $\vec{M}$ is assumed to be uniform over the film and coherently rotated upon sweeping the external magnetic field. The free energy density consists of dipolar ($E_d$), magnetocrystalline ($E_{mc}$), interface ($E_S$), and Zeeman ($E_Z$) energies, whose expressions are given below respectively [29]:

$$E_d = -2\pi M_s^2 \sin^2\theta \tag{A1}$$

$$E_{mc} = K_1 \sin^2\theta + K_2 \sin^4\theta \tag{A2}$$

$$E_s = K_s \sin^2\theta / t_{Co} \tag{A3}$$

$$E_z = -HM_s \cos\theta \tag{A4}$$

where $K_s$ is the interface anisotropy constant, $K_1$ and $K_2$ the second and forth order magnetocrystalline anisotropy constant, respectively, $M_s$ the saturation magnetization, $H$ the applied magnetic field, $t_{Co}$ the Co thickness, $\theta$ the angles between magnetization and $z$-direction. We assume that $K_1, K_2, K_s > 0$, and are all independent of $t_{Co}$. To determine the equilibrium state of $\vec{M}$ at a specific $H$ value, numerical energy minimization is performed on the total energy density:

$$E = -2\pi M_s^2 \sin^2\theta + K_1 \sin^2\theta + K_2 \sin^4\theta + K_s \sin^2\theta / t_{Co} - HM_s \cos\theta \tag{A5}$$

Before proceeding with the simulations as presented in Section III A, we first take a look at all the anisotropy energy terms to have a better understanding of the macro-spin model. For easy treatment, the effective anisotropy $K_{eff}$ is defined phenomenologically as $K_{eff} = K_1 - 2\pi M_s^2 + K_s / t_{Co}$. In this way, the effective anisotropy energy including $E_{mc}$, $E_d$ and $E_s$ can be rewritten as:

$$E_K = K_{eff}\sin^2\theta + K_2\sin^4\theta \qquad (A6)$$

Eq. (A6) can be further rearranged in the form of

$$E_K = K_2[(\sin^2\theta) + \frac{K_{eff}}{2K_2}]^2 - \frac{K_{eff}^2}{4K_2} \qquad (A7)$$

Eq. (A7) is a special case of Eq. (A5) with $H = 0$, which corresponds to the remanent state of $\vec{M}$. Three different equilibrium states can be inferred from Eq. (A7) depending on the values of $K_{eff}$ and $K_2$, which are summarized as follows:

a) $K_{eff} < -2K_2$

In this case, a minimum of $E_K = K_{eff} + K_2$ is obtained at $\theta = \pm\pi/2$. This suggests that in the present case, $\vec{M}$ lies in-plane at remanence. Since $K_1$, $K_2$, $K_s$ are positive constants (based on our earlier assumption), for $K_{eff} = K_1 - 2\pi M_s^2 + K_s/t_{Co}$ to be negative, it requires a large $t_{Co}$. This explains our experimental observations that at large $t_{Co}$, the Pt/Co bilayer favors in-plane magnetic anisotropy (IMA).

b) $-2K_2 \leq K_{eff} \leq 0$

$E_K$ has a minimum of $-\frac{K_{eff}^2}{4K_2}$ at $\theta = \arcsin(\sqrt{-\frac{K_{eff}}{2K_2}})$ or $\pi - \arcsin(\sqrt{-\frac{K_{eff}}{2K_2}})$. This means that the remanent $\vec{M}$ is at an inclined angle with respect to $z$-direction. Such kind of state has been observed experimentally in PMA films within a specific range of film thickness, and is often referred to as "cone state" in macro-spin approximation.

c) $K_{eff} > 0$

The minimum of $E_K$ is obtained at $\theta = 0, \pi$ with a magnitude of 0. This condition is satisfied at small $t_{Co}$, where stable perpendicular magnetic anisotropy (PMA) is achieved in the system.

Based on the above discussion of Eq. (A5), it is clear that with increasing $t_{Co}$, the Co/Pt bilayer experiences a transition from PMA to IMA via the intermedium "cone state". This transition is often referred as the spin reorientation transition (SRT), which is widely observed in ultra-thin Fe, Ni, Co films [21-23]. Moreover, at the critical thickness of SRT, the remanence can serve as a good indicator for the anisotropy state.

a)Email: shenlei@nus.edu.sg (theory)
b)Email: elewuyh@nus.edu.sg (experiment)

[48] J. Hu and R. Wu, Phys. Rev. Lett. **110**, 097202 (2013).

[49] N. Di, J. Kubal, Z. Zeng, J. Greeley, F. Maroun, and P. Allongue, Appl. Phys. Lett. **106**, 122405 (2015).21

**FIGURE CAPTIONS**

FIG. 1. (a) – (c) AHE loops for Ta(1.5)/Pt(3)/Co($t_{Co}$) with $t_{Co}$ = 0.2 – 1.9 nm. Note that the curves in (a) - (c) and (e) are vertically shifted for clarify.

FIG. 2. (a) Schematic of the coordinate system adopted for deriving Eq. (1); (b) Plot of $-HM_s/(2\cos\theta)$ against $2\sin^2\theta$ from AHE loops of the $t_{Co}$ = 1.9 nm (circle) and $t_{Co}$ = 1.8 nm (triangle) samples, and the linear fitting (solid line); (c) Summary of the estimated $K_{eff}$ values with $t_{Co}$ = 1.6 - 1.9 nm; (d) and (e) Fitting of AHE loops for Ta(1.5)/Pt(3)/Co($t_{Co}$) with $t_{Co}$ = 1.7 – 1.9 nm using the normal distribution; (f) Fitting of the AHE loops with $t_{Co}$ = 1.2 – 1.6 nm without consideration of the normal distribution (open square: experimental data, solid-line: fitting results). Note that the curves in (d) – (f) are vertically shifted for clarify.

FIG. 3. (a) and (b) AHE loops for Ta(1.5)/Pt(3)/Co(0.6) with different oxygen pre-exposure dose; (c) AHE loops for Ta(1.5)/Pt(3)/Co(1.9) with different oxygen pre-exposure doses. The topmost curves in (a) and (c) are obtained after annealing. Note that the curves in (a) – (c) are vertically shifted for clarify.

FIG. 4. (a) Comparison of the AHE loops for pristine samples with $t_{Co}$ = 0.4 nm and 1.8 nm, and O exposed samples with $t_{Co}$ = 0.6 nm and $t_{Co}$ = 1.9 nm; (b) Oxygen dose dependence of $M_r/M_s$ ratio for the $t_{Co}$ = 0.6 nm sample; (c) Oxygen dose dependence of $M_r/M_s$ ratio for the $t_{Co}$ = 1.9 nm sample; (d) Comparison of the AHE loops for Ta(1.5)/Pt(3)/Co(0.8) at pristine, pre-exposed, Mg dusting and re-exposed states; (e) Comparison of the AHE loops for Ta(1.5)/Pt(3)/Co(0.6) at pristine, Mg dusting and re-exposed states. Note that the $M_r/M_s$ ratio of the $t_{Co}$ = 0.4, 1.7, 1.8 nm pristine samples are added in (b) and (c) as a reference, and the curves in (a), (d) and (e) are vertically shifted for clarify.

FIG. 5. Schematics for optimized structures used in *ab initio* calculations of (a) oxygen exposed Pt/Co/O; (b) Mg dusted pre-exposed Pt/Co/O/Mg. Note that Pt, Co, O, Mg are represented by blue, pink, red and green balls, respectively. The charge density difference (top view) plotted using an isovalue of 0.012 $e$ Å$^{-3}$ for (c) Pt/Co/O and (d) Pt/Co/O/Mg. The red (blue) region indicates an accumulation (depletion) of electrons. Majority spin (positive) and minority spin (negative) PDOS on the $p$ orbitals of O and $d$ orbitals of Co in (e) and (f) for Pt/Co/O, and in (g) and (h) for Pt/Co/O/Mg, respectively. The zero of energy is set to be Fermi level and a dashed line is added as a guidance to the eye. Inset in (h) is PDOS on the $d$ orbitals of Co in Pt/Co.

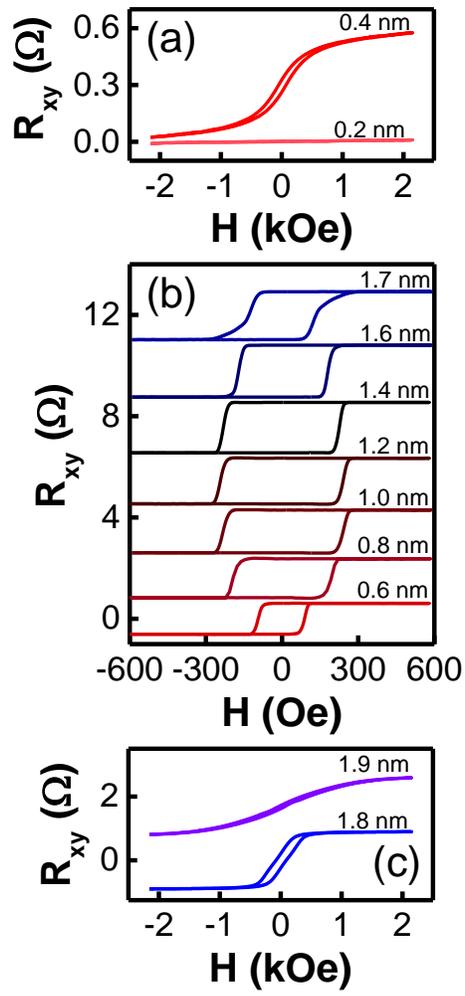

FIG. 1

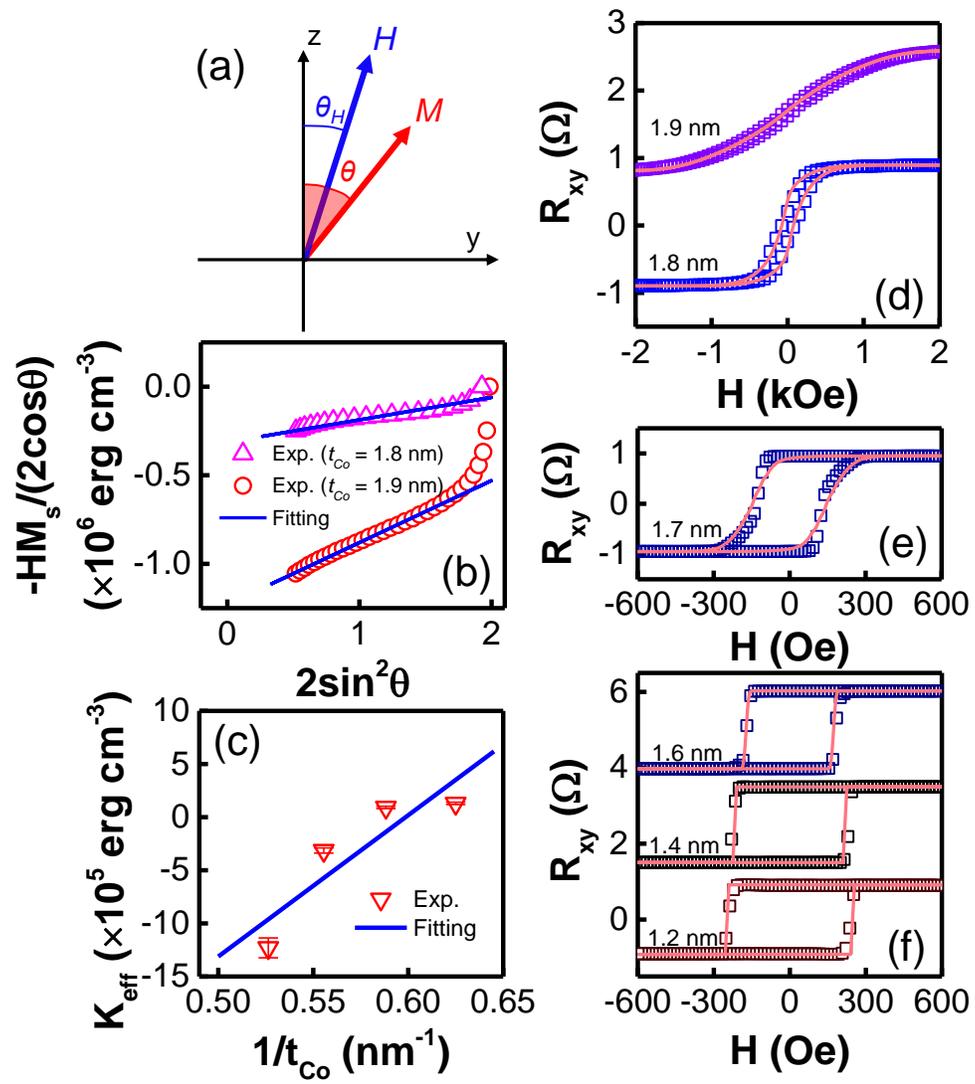

FIG. 2

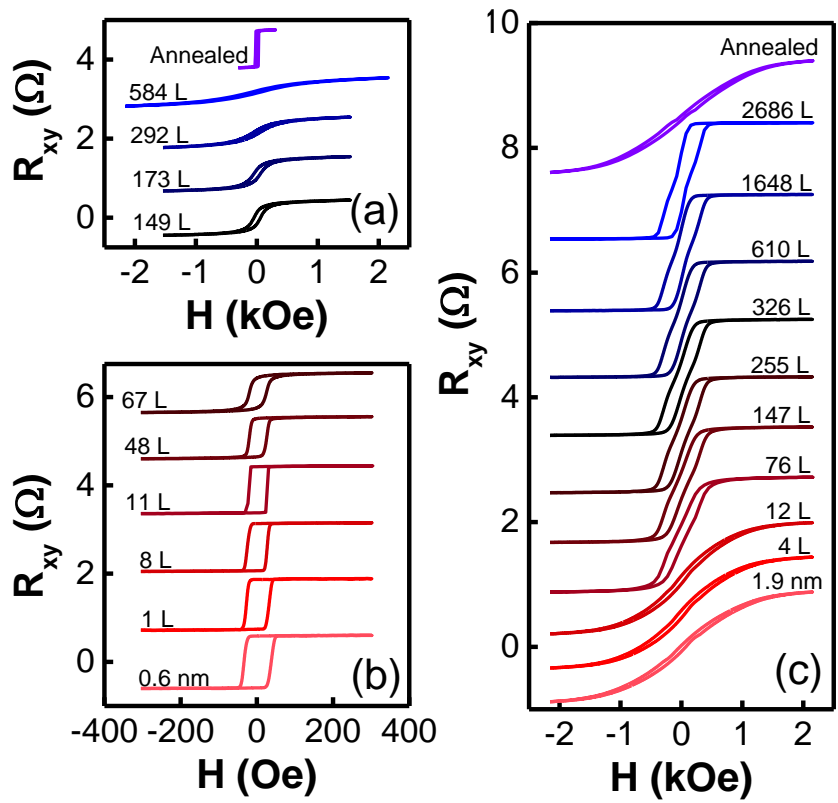

FIG. 3

Physical Review B

Yumeng Yang

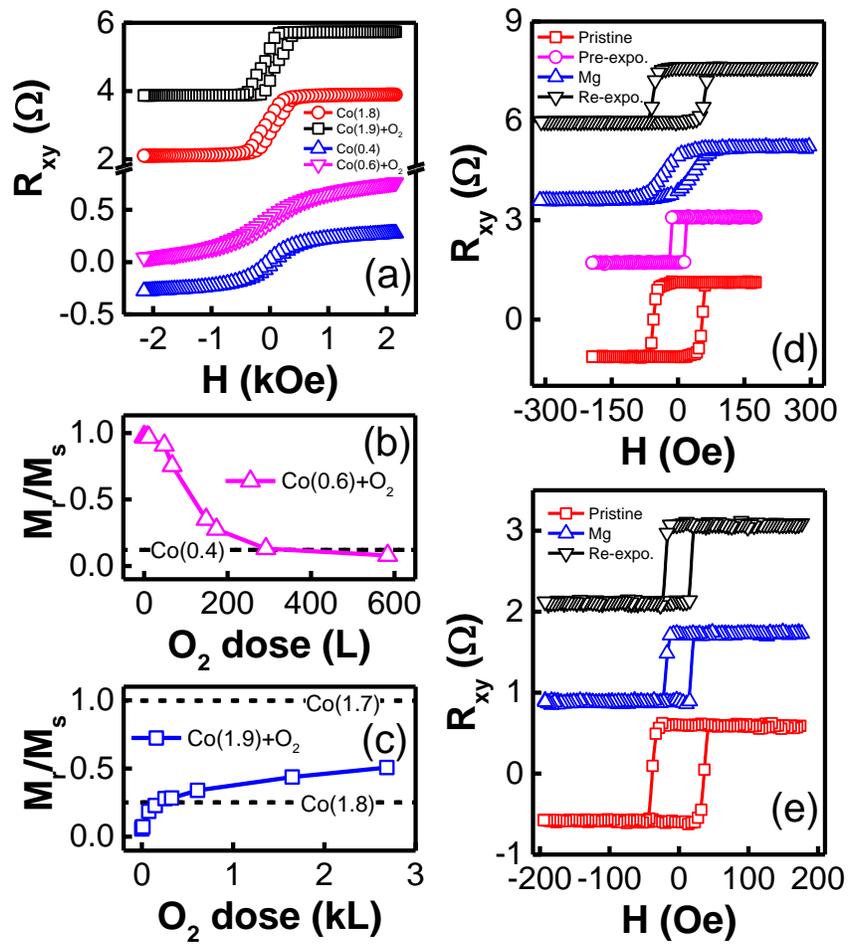

FIG. 4

Physical Review B

Yumeng Yang

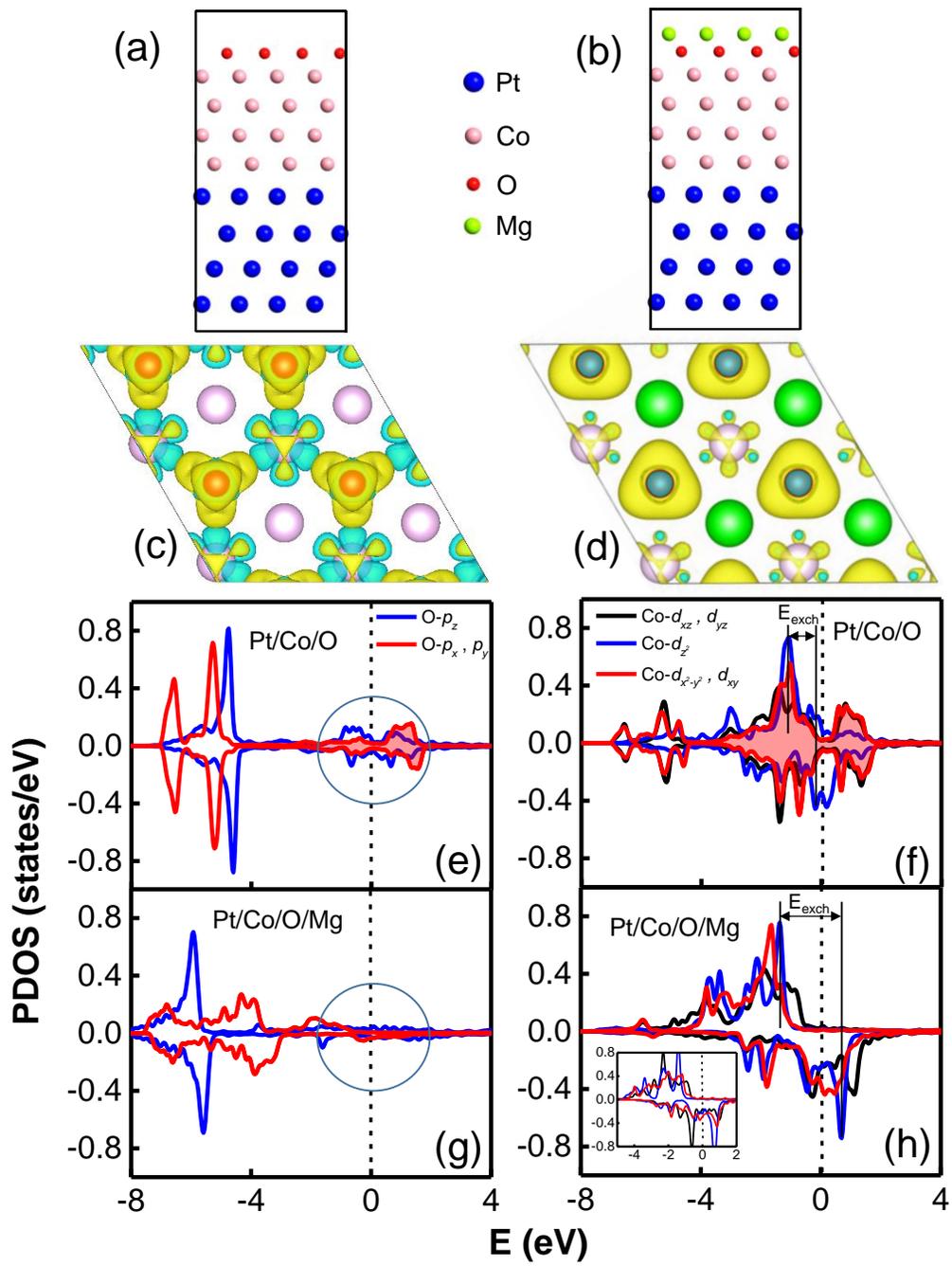

FIG. 5

Table I Summary of the fitting parameters of the AHE loops in the range of $t_{Co}$ = 1.2 - 1.9 nm using Eq. (1).

| $t_{Co}$ (nm) | $K_1$ (erg cm$^{-3}$) | $K_2$ (erg cm$^{-3}$) | $K_s$ (erg cm$^{-2}$) | $\sigma$ (nm) |
|---|---|---|---|---|
| 1.9 | 4.37×10$^6$ | 1.25×10$^5$ | 1.33 | 0.063 |
| 1.8 | 4.85×10$^6$ | 1.20×10$^5$ | 1.33 | 0.060 |
| 1.7 | 4.60×10$^6$ | 1.35×10$^5$ | 1.33 | 0.017 |
| 1.6 | 4.25×10$^6$ | 1.20×10$^5$ | 1.33 | N. A. |
| 1.4 | 3.09×10$^6$ | 1.20×10$^5$ | 1.33 | N. A. |
| 1.2 | 1.53×10$^6$ | 1.20×10$^5$ | 1.33 | N. A. |

Table II Summary of *ab initio* calculated $m_S$, $m_L$ and $K_s$ for the optimized structures of Pt/Co, Pt/Co/O, Pt/Co/O/Mg, respectively.

| Structure | | Moment ($\mu_B$ per atom) | | | | $K_s$ (erg cm$^{-2}$) |
|---|---|---|---|---|---|---|
| | | Co1 | Co2 | Co3 | Co4 | |
| Pt/Co | $m_S$ | 1.83 | 1.72 | 1.74 | 1.82 | 0.71 |
| | $m_L$ | 0.10 | 0.11 | 0.11 | 0.13 | |
| Pt/Co/O | $m_S$ | 1.80 | 1.72 | 1.73 | 0.27 | 0.36 |
| | $m_L$ | 0.10 | 0.11 | 0.13 | 0.01 | |
| Pt/Co/O/Mg | $m_S$ | 1.82 | 1.72 | 1.69 | 1.92 | 0.70 |
| | $m_L$ | 0.10 | 0.11 | 0.12 | 0.13 | |